\documentclass[twocolumn]{jpsj2} %% two-column layout
\usepackage{bm}

\title{Systematic Control of Carrier Doping without Disorder at Interface of Oxide Heterostructures}

\author{Motoaki \textsc{Hirayama}\thanks{E-mail: hirayama@solis.t.u-tokyo.ac.jp} and Masatoshi \textsc{Imada}}

\inst{Department of Applied Physics, University of Tokyo, 7-3-1 Hongo, Bunkyo-ku, Tokyo 113-8656, Japan\\
CREST, JST, 7-3-1 Hongo, Bunkyo-ku, Tokyo 113-8656, Japan}

\abst{ We propose a method to systematically control carrier densities at the interface of transition-metal oxide heterostructures without introducing disorders.
By inserting non-polar layers sandwiched by polar layers, continuous carrier doping into the interface can be realized.
This method enables us to control the total carrier densities per unit cell systematically up to high values of the order unity. 
}

\kword{correlated-electron systems, doping control, heterostructure, interface, two-dimensional electron systems}

\begin{document}
\maketitle

%%%%%%%%%%%%%%%%%%%%%%%%%%%%%%%%%%%%%%%%%%%%%%%%%%%%%%%%%%%%%%%%%%%%%%%%%%%%%%%%%%%%%%%%%%

\section{Introduction}

 Two-dimensional electron systems in semiconductor heterostructures have long been studied from scientific viewpoints as well as from industrial requirements, where carrier concentration can be controlled by the gate voltage \cite{ando}.
In fact, physics of semiconductor developed in the past century has been scientific bases of modern electronics. 
In particular, two-dimensional interfacial microstructures, such as metal oxide semiconductor field effect transistors (MOS-FET) and GaAs heterostructures, have played a major role.
In these systems, fine-control techniques of two-dimensional electrons trapped at the interface have been developed. 
By using these structures, the sheet carrier density has been realized up to about $\sim 10^{13}$cm$^{-2}$ with remarkable suppression of disorder effects.
Quantum Hall effect \cite{stormer} has been observed in this region of low carrier densities under strong magnetic fields.

 Meanwhile, clarification of correlation effects has been one of the most important issues in the condensed matter  physics.
In the electron gas, the correlation effect becomes prominent for smaller density of electrons in general, because the averaged kinetic energy scales as $r_s^{-2}$ for the mean distance $r_s$ between two neighboring electrons, while the mean Coulomb interaction energy scales as $r_s^{-1}$.
However, this rule does not necessarily hold when the electron density approaches a value of filling commensurate with the periodic potential of lattice formed from the atomic nuclei in the crystal.
This commensurate filling can be reached, for example, at the density of one (or simple fractional number) per unit cell for conduction electrons, which is normally the order of $10^{23}$ cm$^{-3}$ in the conventional bulk crystal.
This density is much higher than the density of normal doped semiconductors and the above scaling naively suggests that the correlation effects are negligible in such materials.
However, the commensurability with the lattice leads to completely different physics, where strongly correlated electrons emerge even at such high densities leading to Mott physics and charge ordering phenomena.\cite{Noda}
Indeed, in the strongly-correlated materials with two-dimensional anisotropy such as transition-metal oxides, the on-site Coulomb repulsions strongly dominate physical properties, when the sheet carrier densities reach $10^{14}$-$10^{15}$cm$^{-2}$ and the number of conduction electrons becomes close to a simple integer or fractional number per unit cell on average. 
Examples are found in the cuprate high-$T_{\text{c}}$ superconductors, Mott insulators and anomalous metals.\cite{imada}
The high carrier densities are realized in transition metal oxides conventionally by using chemical doping, which in most cases inevitably introduces disorder, unfortunately. 
Effects from disorder such as Anderson localization \cite{anderson} often obscure intrinsic correlation effects. 
On the other hand, one can realize carrier doping without disorder by FET\cite{ahn}. 
However, carrier density is achieved up to only about $\sim 10^{13}$cm$^{-2}$ by using this method.
If high carrier densities at transition-metal oxides could be realized without disorders, such systems could be ideal experimental stages of low-dimensional strongly-correlated electron systems and would make a major contribution to understanding of the correlation effects. 
\begin{figure}[tb]
\centering
\includegraphics[clip,width=0.45\textwidth ]{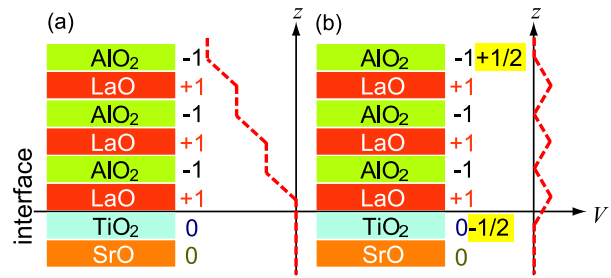} 
\caption{(Color online) Schematic illustrations of reconstruction at interface. Dashed line plots (in red) represent the electronic potential along the [$001$] direction. Integers on the right of the layers represent charges of the layers.  (a) The potential energy diverges along the [$001$] direction. (b) The polar instability can be avoided if TiO$_2$ layer at the interface gets half an electron per unit cell. }
\label{reconstruction_nomal}
\end{figure}

 Transition-metal oxide heterostructures are promising candidates of realizing a disorder-free interface with higher density of carriers within an atomic scale thickness, for example by pulsed laser deposition \cite{kawasaki}.
Recently, Ohtomo and Hwang have reported the existence of a conducting electron layer at TiO$_2$-LaO interface between two band insulators, SrTiO$_3$ and LaAlO$_3$ \cite{ohtomo}.
Other groups also reported the existence of superconductivity at the SrTiO$_3$/LaAlO$_3$ interface\cite{reyren}.

 A mechanism for high conductivity at the transition-metal oxide interfaces has been proposed\cite{nakagawa}.
The key idea is reconstruction of electronic distributions caused by a polar discontinuity, which is essentially understood from classical electromagnetism.
Perovskite structures ($AB$O$_3$) can be divided into alternating layers of $A$O and $B$O$_2$ planes along the [$001$] direction.
Sr$^{2+}$O$^{2-}$ and Ti$^{4+}$O$^{2-} _2$ layers are charge-neutral layers, while in the ionic limit of LaAlO$_3$, La$^{3+}$O$^{2-}$ has a positive charge and Al$^{3+}$O$^{2-} _2$ has a negative charge.    
A polar bilayer, such as La$^{3+}$O$^{2-}$-Al$^{3+}$O$^{2-} _2$, makes no electric-field outside the bilayer, but gives a finite electric potential difference between the plane above and below the bilayer. 
When the thickness of polar layers increases, the potential energy along the [$001$] direction increases, which yields an electrostatic instability (see Fig. \ref{reconstruction_nomal} (a)).
This instability can be avoided if $1/2$ electron per unit cell is doped from the surface into the Ti sites at the interface (Fig. \ref{reconstruction_nomal} (b)).
These doped carriers contribute to electronic conduction.
In this way, conductive interfaces made of transition-metal oxides can be realized, in principle, without disorders.

 This concept, the polar discontinuity, has been well appreciated since early times of studies on semiconductors\cite{baraff, harrison}.
In Ge/GaAs interface, for example, Ge layers are charge-neutral, while GaAs layers have polarity along the [$001$] direction.
To avoid the instability of the electric potential from the polar layer of GaAs, half of the Ge sites at the interface are replaced with Ga (or As) atoms. 
Namely, the reconstruction occurs in the lattice systems on the atomic level.
On the other hand, in the transition-metal oxide heterostructures, the reconstruction is expected in the electron system. 

 Here we note that, this mechanism enables us to dope only a discrete value of the carrier density at the interfaces\cite{thiel, huijiben}.
To investigate the nature of correlation effects, however, continuous tuning of the carrier densities is desired\cite{imada}.

In this paper, we propose a method to control the carrier densities continuously without disorders by a specific procedure of doping into the transition-metal oxide heterostructures.
We demonstrate this charge controllability in the following chapter.

%%%%%%%%%%%%%%%%%%%%%%%%%%%%%%%%%%%%%%%%%%%%%%%%%%%%%%%%%%

\section{Classical Calculation}

\subsection{Doping into Interface}
 In this chapter, we propose a method to systematically control carrier densities doped into the transition-metal oxide heterostructures without disorders. 
This method enables us to change the carrier densities continuously. 
Moreover, this method enables us to realize high densities of carrier doping.
The idea of the systematic doping is very simple.
We insert non-polar layers into the polar layers.
Then, the doped carrier density changes, so that it is optimized to avoid the instability of the potential energy.   
In this paper, we refer to the non-polar perovskite layers inserted into the polar layers as a \textit{$spacer$}.

\begin{figure}[tb]
\centering
\includegraphics[clip,width=0.45\textwidth ]{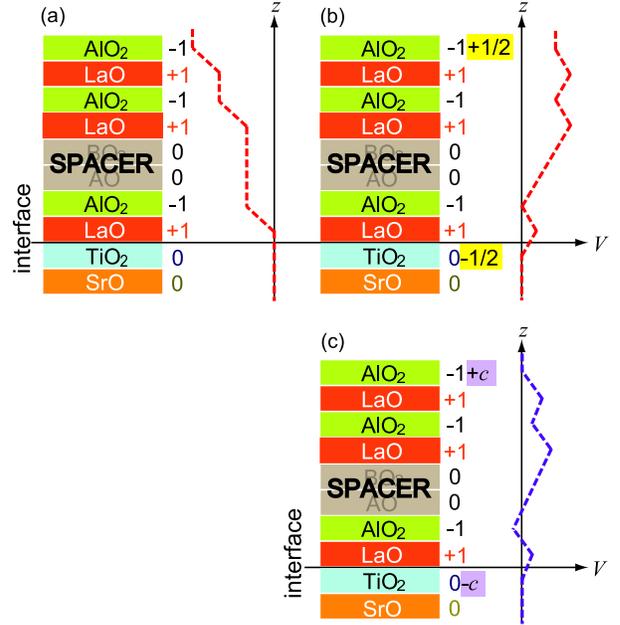} 
\caption{(Color online) Schematic illustrations of reconstruction at interface with non-polar spacer. The (red and purple) dashed lines represent the electronic potential along the [$001$] direction. Integers on the right of the layers represent charges of the layers.  (a) When a non-polar spacer is inserted into the polar layer, the potential difference between the interface and the surface does not change compared to the case of Fig. \ref{reconstruction_nomal} (a) within the interval of the spacer, if there is no reconstruction. (b) On the other hand, in the case where half an electron is doped to the interface, the divergence of the electric potential is only incompletely avoided compared to the case of Fig. \ref{reconstruction_nomal} (b). (c) Half doping into the interface is no longer the best way to avoid the divergence, and the optimum amount of the carrier filling changes from $1/2$ to another value $c$.}
\label{reconstruction}
\end{figure}

 An essence of the method can be understood in classical physics. 
Let us consider non-polar perovskite, which has wide band gap and a lattice spacing close to the transition-metal oxide interface, as the spacer.
Figure \ref{reconstruction} shows an interface where a spacer is inserted above a certain AlO$_2$ layer.
The electric potential along the [$001$] direction does not increase in the spacer region, if a reconstruction does not occur (see Fig. \ref{reconstruction} (a)).
The potential difference between the interface and the surface of the heterostructure is obviously the same in comparison with the non-inserted one (Fig. \ref{reconstruction_nomal} (a)).
However, in case that half an electron per unit cell is doped into the interface, the instability of the potential divergence along the [$001$] direction is only incompletely removed, because the electric potential increases in the region of the spacer (Fig. \ref{reconstruction} (b)). 
In this case, doping $1/2$ electron per unit cell into the interface is no longer the best way to avoid the potential divergence, and the optimum amount of carrier density changes from $1/2$ to another value (Fig. \ref{reconstruction} (c)).

Now, we define $L$ as the number of the polar layers, $m_{B\text{O}_2}$ as the number of the spacer inserted above $B$O$_2$ (e.g. Fig. \ref{reconstruction} (a)), and $m_{A\text{O}}$ as the number of spacer inserted above $A$O of the polar layer.
We define a single layer (or a single spacer) as one unit cell along the [$001$] direction. 
The total energy per unit cell is obtained in the classical electromagnetism as  
\begin{equation}
E=(L+m_{A\text{O}}+m_{B\text{O}_2})c^2-(L+2m _{A\text{O}})c\times 1+\text{const.} ,
\label{eq:E}
\end{equation}
where $c$ is the doped carrier density per unit cell at the interface.
In this section, we employ the energy unit by $1/2\epsilon$,  where $\epsilon$ is the dielectricity, for the sake of simplicity.
Differentiating eq. (\ref{eq:E}) with respect to the carrier density $c$, we obtain the optimum number of the carrier density as
\begin{equation}
c=\frac{1}{2}\times \frac{L+2m _{A\text{O}}}{L+m_{A\text{O}}+m_{B\text{O}_2}}.
\label{eq:c}
\end{equation}
This result indicates that the carrier density can be controlled from $0$ to $1$ by tuning the ratio of $m_{B\text{O}_2}$ and $m_{A\text{O}}$ to $L$.  
Instead of the non-polar layers, if we can use materials with other polarities as spacers, we obtain 
\begin{equation}
c=\frac{1}{2}\times \frac{(L+2m_{A\text{O}})a+(-m_{A\text{O}}+m_{B\text{O}_2})b}{L+m_{B\text{O}_2}+m_{A\text{O}}},
\label{eq:ce}
\end{equation}
where $a$ is the polarized charge of the polar bilayer ($+a$, $-a$) and $b$ is the charge of the spacer ($+b$, $-b$).
Equation (\ref{eq:c}) represents the limit of the perfect polarity of the polar layer, $a\rightarrow 1$ and the perfect non-polarity of the spacer, $b\rightarrow 0$.
In addition, if we consider the dissociation energy of an oxygen from the surface $E_s$, which is the energy cost to generate the carrier at the surface, then the total energy is obtained as
\begin{equation}
E=(L+m_{A\text{O}}+m_{B\text{O}_2})c^2-(L+2m _{A\text{O}})c\times 1+E_s c +\text{const.} .
\label{eq:Eve}
\end{equation}
Differentiating eq. (\ref{eq:Eve}) with respect to $c$, we obtain the optimum number of the carrier density as
\begin{equation}
c=\min \left(\frac{1}{2}\times \frac{L+2m _{A\text{O}} -E_s}{L+m_{A\text{O}}+m_{B\text{O}_2}},0\right).
\label{eq:cve}
\end{equation}
This optimum number $c$ decreases from $1/2$ (in the case without the spacer) because of the energy cost term of $E_s$.
The other factors, the kinetic term, the hybridization effect at the interface, the lattice relaxation, and the atomic exchange, also decrease the carrier density at the interface. 
These effects will be discussed in the last section.

%%%%%%%%%%%%%%%%%%%%%%%%%%%%%%%%%%%%%%%%%%%%%%%%%%%%%%%%%%%%%%%%%%%%%%%%%%%%
 
\subsection{Doping into both Interface and Spacer}
\begin{figure}[tb]
\centering
\includegraphics[clip,width=0.3\textwidth ]{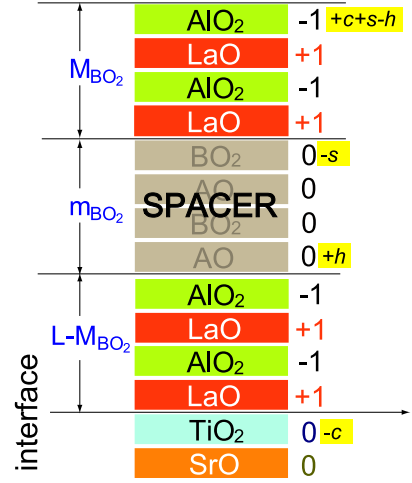} 
\caption{(Color online) Schematic of our model. Doped carriers are confined near the substrate (with the concentration $-c$) and near the spacer region (with the concentration $-s$). We also consider hole doping (with the concentration $h$) at the $p$-type interface between the spacer and the polar layer.}
\label{DtS}
\end{figure}

Next, to consider the effects of doping into the spacer in detail, we qualitatively estimate the optimum amount of the carrier in the substrate and spacer region again in the classical limit. 
We consider the case where the only one spacer is inserted above $B$O$_2$ layer.
The total energy of the interface mainly consists of $3$ terms, the electric interaction term, the energy level of sites, and the dislocation energy.
The electrostatic energies can be calculated as a capacitor, where the energy is proportional to the thickness of the capacitor and the square of accumulated charge.
We consider the hole doping into a $p$-type interface as well as the electron doping into the $n$-type interface (see Fig. \ref{DtS}).
Then the total energy is given by
\begin{multline}
E=(L+m_{B\text{O}_2})c^2 + M_{B\text{O}_2}s^2 +(m_{B\text{O}_2} +M_{B\text{O}_2})h^2\\
  + 2M_{B\text{O}_2} s c -2(m_{B\text{O}_2}+M_{B\text{O}_2})ch-2M_{B\text{O}_2}sh \\
  -L c -M_{B\text{O}_2}s +M_{B\text{O}_2}h +Vs +V'h +E_s(c+s-h)+\text{const.} ,
\label{Espeh}
\end{multline}
where $M_{B\text{O}_2}$ is the number of the polar layers above of the spacer, $V$ and $V'$ are the energy levels of the sites in the spacer region measured from the level of the substrate, $s$ is the doped carrier density at the $n$-type interface of spacer, and $h$ is the hole density at the $p$-type interface of spacer.
To minimize the total energy eq. (\ref{Espeh}), the optimum carrier densities are obtained as
\begin{equation}
c=\max \left(\frac{1}{2}\times \frac{L-M_{B\text{O}_2}-V'}{L-M_{B\text{O}_2}},0\right) ,
\label{cspeh}
\end{equation}
\begin{equation}
s=\max \left(\frac{1}{2}\times \frac{M_{B\text{O}_2}-V-E_s}{M_{B\text{O}_2}} -\frac{1}{2}\times \frac{V+V'}{m_{B\text{O}_2}},0\right)
\label{sspeh}
\end{equation}
and
\begin{equation}
h=\max \left(\frac{1}{2}\times \frac{L-M_{B\text{O}_2}-V'}{L-M_{B\text{O}_2}}-\frac{1}{2}\times \frac{V+V'}{m_{B\text{O}_2}},0\right),
\label{hspeh}
\end{equation}
where $c$, $s$, and $h$ are $\geq 0$.
We find that $c$ and $h$ in eqs. (\ref{cspeh}) and (\ref{hspeh}) do not depend on $E_s$. 
If $E_s$ increases, the doped carrier $s$ in the spacer region decreases first, and next, when $s$ vanishes, the doped carrier $c$ and hole $h$ starts decreasing.
A small change in $E_s$ has no effect on the carrier density at the interface.
To consider the case $h=0$, a threshold $V'_{\text{th}}$ exists for the hole doing as
\begin{equation}
V'_{\text{th}} =\frac{(m_{B\text{O}_2}-V)(L-M_{B\text{O}_2})}{L-M_{B\text{O}_2}+m_{B\text{O}_2}}.
\label{shV}
\end{equation}
If $V' < V'_{\text{th}} $ is satisfied, the holes are not doped at the spacer.
In this case, the optimum carrier densities are obtained as 
\begin{equation}
c=\max \left(\frac{1}{2}\times \frac{L-M_{B\text{O}_2}+V}{L+m_{B\text{O}_2}-M_{B\text{O}_2}},0\right)
\label{cspe}
\end{equation}
and 
\begin{align}
s&=\max \biggl(-\frac{1}{2}\times \frac{L-M_{B\text{O}_2}+V}{L+m_{B\text{O}_2}-M_{B\text{O}_2}}\notag\\
          &+\frac{1}{2}\times \frac{M_{B\text{O}_2}-V-E_s}{2M_{B\text{O}_2}}, 0\biggr) .
\label{sspe}
\end{align}
In eq. (\ref{cspe}), $c$ does not depend on $E_s$ again.

%%%%%%%%%%%%%%%%%%%%%%%%%%%%%%%%%%%%%%%%%%%%%%%%%%%%%%%
 
\section{Quantum Calculation by Hartree Approximation}  
 
\subsection{Formalism} 

\begin{figure}[tb]
\centering
\includegraphics[clip,width=0.3\textwidth ]{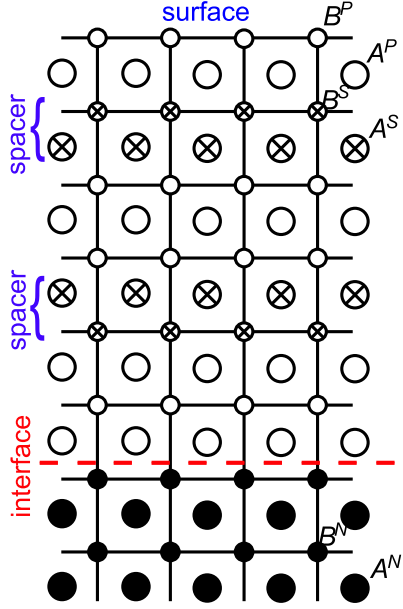} 
\caption{(Color online) Schematic of our model. Filled circles represent cations of non-polar substrate $A^N B^N$O$_3$, open circles represent cations of polar substrate $A^P B^P$O$_3$ and cross marked circles represent cations of non-polar spacer $A^S B^S$O$_3$. The doped electrons are assumed to transfer between the $B$ sites in our model.}
\label{model}
\end{figure}

 Next, we discuss quantum effects, namely, the effects of itinerancy of electrons, in our model.
The carriers are expected to extend over several layers by the quantum effects.     
Moreover, if we consider the energy levels of the sites at the interface, the carriers are expected to be doped not only into the interface region, but also into the spacer region. 

 Figure \ref{model} illustrates a heterostructure which we consider in this paper.
As shown in Fig. \ref{model}, $A^P B^P$O$_3$ has a polarity, while the substrate $A^N B^N$O$_3$ and the spacer $A^S B^S$O$_3$ do not.
Each layer is a band insulator in the bulk.
We neglect the orbital degrees of freedom of the transition metal $d$ orbitals.
We also neglect charge fluctuations of the valence bands of the oxygen $2p$ orbitals in this calculation.
Our simplified Hamiltonian of this system is written as
\begin{equation}
\mathcal{H}=\mathcal{H}_{\text{hop}}+\sum_{i}\mathcal{H}^i _{\text{pot}}+\mathcal{H}_{\text{surface}},
\label{Hamiltonian}
\end{equation}
with $i$ being the site index.
The hopping term $\mathcal{H}_{\text{hop}}$ is defined as  
\begin{equation}
\mathcal{H}_{\text{hop}}=-\sum_{\langle ij\rangle ,\sigma }t_{ij}(c_{i\sigma }^{\dagger }c_{j\sigma} +c_{j\sigma }^{\dagger }c_{i\sigma}),
\end{equation}
where $c_{i\sigma }^{\dagger }$ ($c_{i\sigma }$) is the creation (annihilation) operator of an electron on the $i$th-site with spin $\sigma$.
The summation over the nearest-neighbor sites is represented by $\langle ij\rangle $. 
The potential term $\mathcal{H}^i _{\text{pot}}$ consists of three terms;
\begin{equation}
\mathcal{H}^i _{\text{pot}}=\mathcal{H}^{i}_{\text{site}}+\mathcal{H}^{i}_{\text{Coulomb}}+\mathcal{H}^{i}_{\text{on-site}}.
\end{equation}
The one-body energy level is defined as 
\begin{equation}
\mathcal{H}^{i}_{\text{site}}=V_{i}n_i ,
\end{equation}
where $n_i=\sum_{\sigma }c_{i\sigma }^{\dagger }c_{i\sigma}$ is the number operator of the doped electrons.

The long-range Coulomb interaction is defined as
\begin{align}
\mathcal{H}^{i}_{\text{Coulomb}} =-\sum_{j\in A^P}\frac{e^2n_i}{\epsilon |\bm{R}^{A^P} _{j}-\bm{r}_{i}|}
+\sum_{j\in B^P}\frac{e^2n_i}{\epsilon |\bm{R}^{B^P} _{j}-\bm{r}_{i}|} \nonumber \\  
+ \sum_{j\in \text{O}^{\text{defect}}}\frac{2e^2n_i}{\epsilon |\bm{R}^{\text{O}^{\text{defect}}} _{j}-\Vec{\bm{r}}_{i}|}    
+\frac{1}{2}\sum_{j\neq i}
   \frac{e^2 n_j n_i}{\epsilon |\bm{r}_{j}-\bm{r}_{i}|},
\end{align}
where the first and second terms are the Coulomb potential from the polar layer.
Here, $A^P$ and $B^P$ represent the $A$ sites and the $B$ sites of the polar layers $A^P B^B$O$_3$ respectively, and the third term comes from defects of oxygen.
The last term is Coulomb repulsion between the doped electrons.  
For simplicity, the dielectric constants are assumed to be uniform in all the regions of the heterostructure and the charge of the polarized layers are assumed to be $\pm e$ while an oxygen defect is assumed to generate the charge $-2 e$.
The on-site Coulomb potential term is defined as 
\begin{equation}
\mathcal{H}^{i}_{\text{on-site}}=U_{i}n_{i\uparrow }n_{i\downarrow }.
\label{onsite}
\end{equation} 
The reconstruction energy of the surface is expressed by  
\begin{equation}
\mathcal{H}_{\text{surface}}=\sum_{i\in \text{O}^{\text{defect}}}\frac{1}{2}E_{s}(i) n^{\text{O}^{\text{defect}}}_{i}, 
\label{sur}
\end{equation}
where $E_s(i)$ is the dissociation energy of an oxygen from the surface.
In this study, we assume that the oxygens dissociate only from the single layer at the surface.
From the electroneutrality condition, the carrier density and the oxygen-defect density satisfy the following condition: 
\begin{equation}
\sum_{i\sigma}n_{i\sigma}=\frac{1}{2}\sum_{i} n^{\text{O}^{\text{defect}}}_{i}. 
\end{equation}
Thus,  eq. (\ref{sur}) can be written as  
\begin{equation}
\mathcal{H}_{\text{surface}}=\frac{1}{2}E_{s}\sum_{i\in \text{O}^{\text{defect}}}n^{\text{O}^{\text{defect}}}_{i}=E_{s}\sum_{i\sigma}n_{i\sigma},
\end{equation}
where $E_s$ has a unique value in each material.

To consider the quantum effects quantitatively, we use the following parameters in our calculation.  
Lattice parameter $a$ is fixed at $3.9$\AA, which corresponds to the experimental lattice parameter of the SrTiO$_3$ substrate \cite{ohtomo}.  
The hopping parameter $t$ is set to $0.3$ eV estimated from the bulk materials \cite{okamoto}.
We take on-site Coulomb interaction in the substrate $U_{\text{substrate}}= 6$ eV ($=18t$) estimated from high energy spectroscopies \cite{saitoh}.
We employ the dielectric constant $\epsilon = 15$ and dimensionless parameter $e^2/\epsilon \cdot a \cdot t = 0.8$.
The effect of the dielectric constant will be discussed in the last section.
The energy level in polar layer $V_{\text{polar}}$ is assumed to be $4.5$ eV ($=15t$) as is estimated from the bulk LaAlO$_3$ band structure\cite{lim}.
We neglect the on-site Coulomb interaction in the polar layer ($U_{\text{polar}}=0$ eV) because the Al $3s$ wave function in LaAlO$_3$/SrTiO$_3$ is widely extended compared to $d$ and $f$ orbitals.
We leave the value of the dissociation energy $E_s$ as a parameter and examine the $E_s$ dependence in the range $0\sim 5$ eV.
Different choices of $E_s$ only affect the carrier density quantitatively and make no change in the essential mechanism of doping, if the doped carrier $c$ is positive.  
The on-site Coulomb interaction in the spacer, $U_{\text{spacer}}$ is also ignored for the sake of simplicity.
To study the properties of this model, we employ the Hartree approximation: $n_{i\uparrow }n_{i\downarrow }$ $\sim$  $\langle n_{i\uparrow } \rangle n_{i\downarrow } +n_{i\uparrow } \langle n_{i\downarrow } \rangle -\langle n_{i\uparrow } \rangle \langle n_{i\downarrow } \rangle$. 
The results for the $100 \times 100 \times 50$ site will be shown below.

%%%%%%%%%%%%%%%%%%%%%%%%%%%%%%%%%%%%%%%%%%%%%%%%%%%%%%%%%%%%%%%%%

\subsection{Quantum Calculation without Spacers}

\begin{figure}[tb]
\centering
\includegraphics[clip,width=0.45\textwidth ]{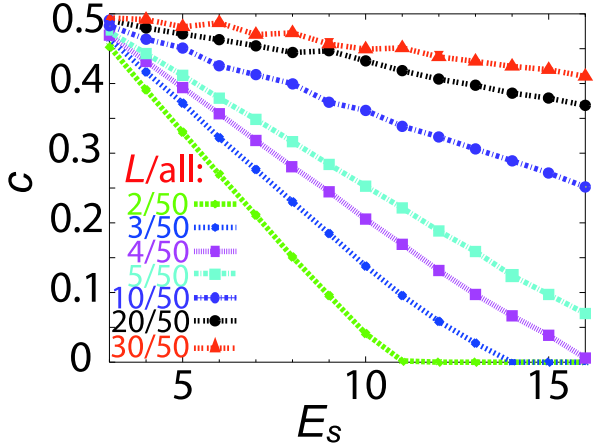} 
\caption{(Color online) Total carrier density $c$ as a function of dissociation energy $E_s$. Total number of layers is taken as $50$. Number of the polar layers $L$ is indicated in the figure. By increasing the number of polar layers, the doped carrier densities increase. Increase in $E_s$ leads to decrease in the carrier density from $c=0.5$.}
\label{non-spacer_Es}
\end{figure}

\begin{figure}[tb]
\centering
\includegraphics[clip,width=0.45\textwidth ]{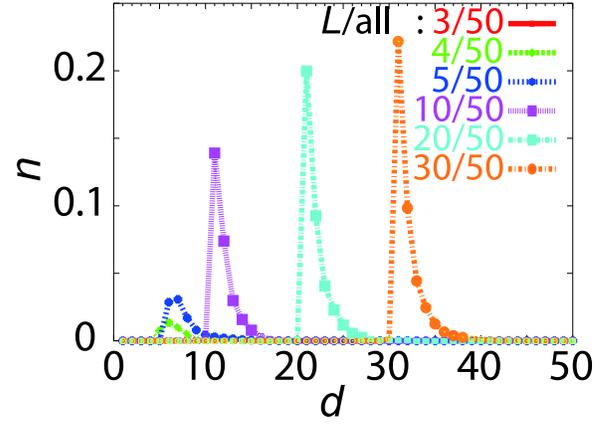} 
\caption{(Color online) Carrier density $n$ as a function of distance from surface $d$. Total number of layers is fixed at $50$. Number of the polar layers $L$ deposited between the surface and the interface is indicated in the figure as $L/50$. Dissociation energy $E_s$ is fixed at $15t$. The doped carriers are confined around the interface layer in the side of non-polar layers.}
\label{non-spacer_dis}
\end{figure}

 We first calculate the charge distribution at the interface in the case without the non-polar spacer.
The carrier densities of the interfaces are determined mainly by the balance between the potential energy and the dissociation energy of the oxygen at the surface. 
As the thickness of the polar layers increases, the carrier densities tend to increase (see Fig. \ref{non-spacer_Es}). 
This tendency is indeed found in real materials \cite{thiel, huijiben}.
Increase in the dissociation energy $E_s$ leads to decrease in the carrier density from $c=0.5$.

 As shown in Fig. \ref{non-spacer_dis}, the doped carriers tend to be confined near the interface.
This tendency is also found in real materials \cite{nakagawa, willmott}.
The polar instability is essentially determined by $L$, $E_s$, $\epsilon$, and $V_i$, whereas the spread of the doped carrier extended away from the interface does neither alter the instability nor the charge $c$.
It only relaxes the local energy (we will discuss these effects in detail in the last section).

%%%%%%%%%%%%%%%%%%%%%%%%%%%%%%%%%%%%%%%%%%%%%%%%%%%%%%%%%%%%%%%%%%%%%%%%%%%%%%%

\subsection{Quantum Calculation with Spacer }

\begin{figure}[tb]
\centering
\includegraphics[clip,width=0.45\textwidth ]{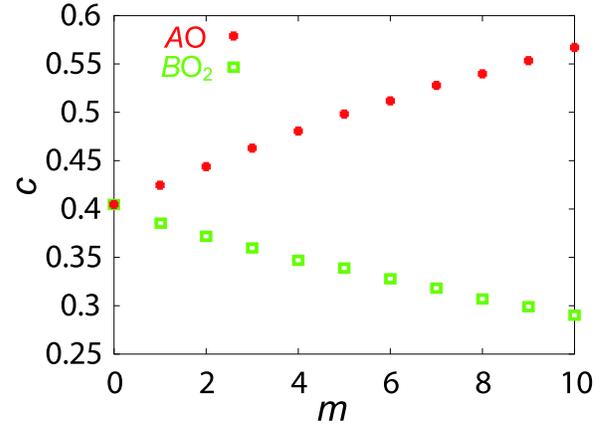} 
\caption{(Color online) Total carrier density $c$ vs. number of spacer $m$. Energy level $V_{\text{spacer}}$ is fixed at $15t$. Here, $E_s$ is assumed to be $15t$. Total number of layers is assumed to be $50$ and the number of the polar layer is assumed to be $25$. The spacers are inserted either between the $B$O$_2$ and $A$O layers in the fifteenth polar layer or on the top of the fifteenth polar layer from the surface . Circle (red) plots show the result of the former case and open (green) squares show the latter.}
\label{c25}
\end{figure}

We now calculate the charge distribution at the interface in the case with spacers.
The calculation was performed in a way similar to that in the previous subsection.
Figure \ref{c25} shows the doped carrier density as a function of thickness of the spacer.
The spacer is inserted either between the $B$O$_2$ and $A$O layers in the fifteenth polar layer (contributing to $m_{A\text{O}}$), or on the top of the fifteenth polar layer counted from the surface (contributing to $m_{B\text{O}_2}$).
In the former case, the insertion of only $m_{A\text{O}}=10$ spacers into $25$ polar layers increase the carrier density up to around $50$ percent compared to the non-inserted case.
Furthermore, by the other type of spacers with $m_{B\text{O}_2}$, the carrier density may not only be increased but also be decreased.
In this way, the carrier density can be finely tuned similarly to the classical estimate in eq. (\ref{eq:cve}).

\begin{figure}[tb]
\centering
\includegraphics[clip,width=0.45\textwidth ]{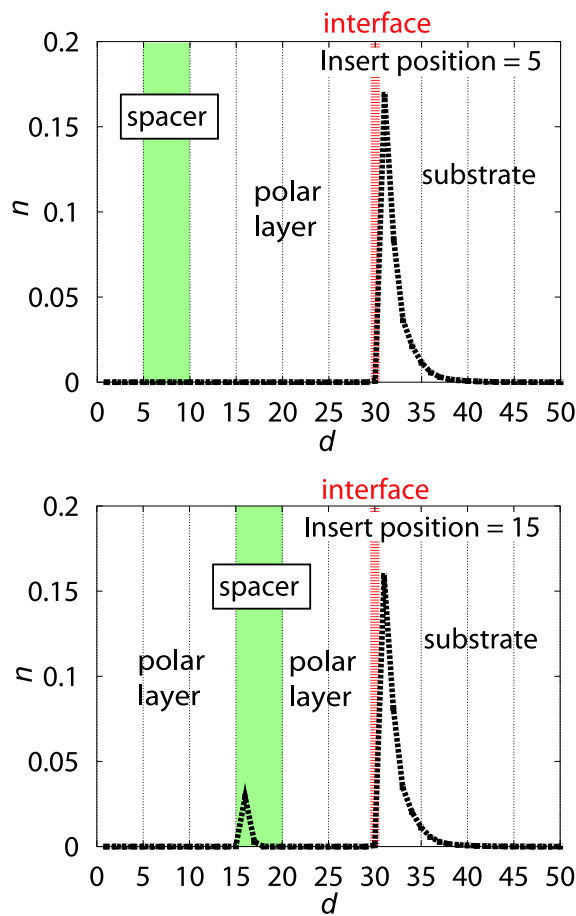} 
\caption{(Color online) Spacer-position dependence of carrier density $n$. Here, $V_{\text{spacer}}$ is fixed at $0$ eV. Dissociation energy $E_s$ is fixed at $15t$. The number of total layers, polar layers and the thickness of spacer are taken to be $50$, $25$, and $5$, respectively. The spacers are inserted on the top of $B$O$_2$ layer of the polar layer. ``Insert position'' specifies the position of spacer counted from the surface.}
\label{dope-even2}
\end{figure}

 Next, we study the dependence of the doped carrier densities on the positions of the spacer inserted. 
Here, as an example, the spacer is inserted above the $B$O$_2$ layer, and the number of polar layers and the thickness of spacer are fixed at $25$ and $5$, respectively.
The energy level $V_{\text{spacer}}$ is fixed at $0$ eV. 
Figure \ref{dope-even2} shows how the carrier density depends on the inserted position of the spacer.
 
A characteristic feature is that, when the spacer is inserted far from the surface, the carrier doping occurs in the spacer region as well.
Whether the carriers are partially doped into the spacer region is determined by the position of the spacer. 
If the spacer is placed farther from the surface, the thickness of the polar layer between the spacer and the surface becomes larger and  lowering of the energy by the carrier doping into the spacer region also becomes larger (see Fig. \ref{dope-even2}).
As a result, carriers are doped partially into the spacer region.
Of course, in the limit where the energy levels of the sites in the spacer region is sufficiently high, the carrier doping occurs only around the normal interface region. 
In this calculation, we neglect roles of the oxygen valence bands in the spacer because their energy levels are too low.

In contrast to the electron doping, the lowering of energy by the hole doping into the spacer region is enhanced when the spacer is not located near the interface but near the surface.            
This tendency is in contrast to the electron doping into the spacer.
With this opposite tendency kept in mind, we should adjust the position of the spacer in order to control the carrier doping into the spacer.   

%%%%%%%%%%%%%%%%%%%%%%%%%%%%%%%%%%%%%%%%%%%%%%%%%%%%%%%%%%%%%%%%%%%%%%%%%%%%%%%

\section{Why are Carrier Densities so low at the real Transition-Metal Oxide Interfaces?}

\begin{figure}[tb]
\centering
\includegraphics[clip,width=0.4\textwidth ]{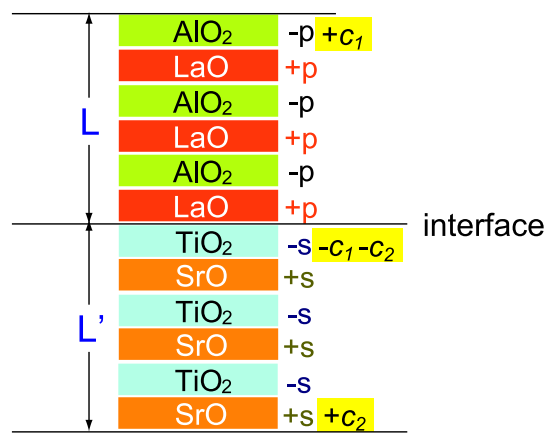} 
\caption{ Schematic of heterostructure for the case deviated from the ionic limit. Here, $\pm p$ and $\pm s$ represent charges of the layers. Carriers at the interface are doped not only from the surface of the polar layer (with the concentration $-c_1$) but also from the bottom of the substrate (with the concentration $-c_2$).  }
\label{polar_doubleside}
\end{figure}

In this section, we examine the reasons why carrier densities are much lower than the expectation from the polar discontinuity in the available experimental conditions at the real transition-metal oxide interfaces.
Later, we will also show that our doping method is consistent even in such low density cases seen in the experiments.   
 
Several works have found carrier densities of $n$-type interfaces around $10^{13}$ cm$^{-2}$ much lower than the prediction considered in this paper, when the density of oxygen vacancies is expected to be low\cite{eckstein}. 
In fact, this is more than one order of magnitude less than $0.5$ electron per unit cell derived in this paper.

There are mainly three reasons for the low carrier densities at the interface.
One of them is the insufficient thickness of the polar layers of the heterostructures.
We have shown the thickness dependence of the doped carrier densities in the previous section.  
In the interface without the spacers, the doped carrier density $c$ is obtained as  
\begin{equation}
c=\max \left(\frac{1}{2}\times (1-\frac{\epsilon E_{s}}{L}),0\right) ,
\end{equation}
where $\epsilon $ is the dielectric constant of the polar layers, $E_{s}$ is the energy cost of reconstruction and $L$ is the number of the polar layers.
If $L$ is small compared to $\epsilon $ and $E_{s}$, the carrier density $c$ can be low or zero.
This thickness dependence of carrier densities is negligible in the limit of sufficiently thick $L\rightarrow \infty $.  
   
Even in such a thick limit, defects and substitutions of atoms may still have influences on the carrier densities at the interfaces.        
In the case where the interface has additional charges of electronic carriers $c$ and atoms $c'$, the total energy is obtained as
\begin{equation}
E=L(c+c')^2-L(c+c')\times 1+\text{const.} ,
\label{eqce}
\end{equation}
and the optimized carrier density is obtained as
\begin{equation}
c=\frac{1}{2}-c'.
\label{cce}
\end{equation}
The instability of the potential divergence is removed by the reconstruction of both the electron and lattice systems (see eq. (\ref{eqce})).   
If the most part of the reconstruction takes place in the lattice system, electronic carriers doped into the interface may become few.
 
From the perspective of the polar discontinuity, the deviation from the ionic limit of the substrate also affects the expectation value of the doped carrier at the interface.  
In the ionic limit, the total energy is obtained in the classical electromagnetism as   
\begin{equation}
E=L c^2-Lc+\text{const.} ,
\label{Eionic}
\end{equation}
where $c$ is the doped carrier density per unit cell at the interface and $L$ is the thickness of the polar layer along the [$001$] direction. 
Differentiating eq. (\ref{Eionic}) with respect to the carrier density $c$, we obtain the optimum number of the carrier density as $c=1/2$.
On the other hand, if the system is away from the ionic limit, the total energy is obtained as
\begin{equation}
E=L c _1 ^2 +L' c _2 ^2 -L p c_1 +L' s c_2 + \text{const.},
\label{Edevi}
\end{equation}
where  $p$ is the charge of the polar bilayer ($+p$, $-p$), $s$ is the charge of the ``polarized'' substrate bilayer ($+s$, $-s$),  $c_1$ and $c_2$ are the amounts of the charge at the surface of the polar layer and the substrate, respectively, and $L'$ is the thickness of the substrate along the [$001$] direction (see Fig. (\ref{polar_doubleside})).
The carrier density at the interface, $c$ is obtained as the sum of $c_1$ and $c_2$.
Differentiating eq. (\ref{Edevi}) with respect to the carrier density $c_1$ and $c_2$, we obtain the optimum carrier density as
\begin{equation}
c=\frac{1}{2} (p - s).
\label{cdevi}
\end{equation}
We estimate charge of the LaO-AlO$_2$ in the bulk LaAlO$_3$ and SrO-TiO$_2$ in the bulk SrTiO$_3$ by the local density approximation.
The calculations were carried out with an in-house code, which is based on the Full-Potential Linear Muffin-Tin Orbitals (FP-LMTO) method\cite{methfessel}.
The lattice parameters are fixed at 3.905 \AA , which corresponds to the experimental lattice parameter of the bulk SrTiO$_3$.
In our calculations, LaO-AlO$_2$ has the charge distribution ($+1$, $-1$) and SrO-TiO$_2$ has ($+0.2$, $-0.2$).
Similar results of the charge distributions have been calculated by the generalized gradient approximation (GGA) \cite{ishibashi}.  
Using these results, $c$ is expected to be reduced by $20$ percent from $1/2$ even in the thick and clean limit.
The strains from the substrates also affect the carrier densities through the change in the ionicity of the polar layers.

These three factors, the thickness, the lattice defects and the deviation from the ionic limit, mainly affect the doped carrier densities at the interfaces.
Other factors at the interfaces, atomic exchange, hybridization, carrier distribution and lattice relaxation, can also reduce total energy of the system and affect the doped carrier densities \cite{ishibashi}.
However, in the limit where the thickness of the polar layers is sufficiently large ($L \rightarrow \infty $), these local effects do not alter the total accumulated charge around the interface.
For example, the lattice relaxation at the interface changes the local polarity and the hybridization between the sites.
However, the potential divergence along the [$001$] direction decreases only near the interface. 
The system in any case requires additional charges, the doped carriers or the defects, to avoid the instability of the potential energy. 
Quantitative explanation for the substantially small concentration of carriers observed experimentally may be given by a combination of the three complex factors together with insufficient thickness of the LAO layers in the available experimental conditions.  

Our method of carrier control depends on the ratio of the thickness of the spacer to that of the polar layer.  
Even if the carrier densities decrease by the defects and the deviation from the ionic limit, the reduced carrier densities are still determined by the thickness of the polar layers and spacers, and our method can be used for the control of the carrier densities at the interfaces.  

%%%%%%%%%%%%%%%%%%%%%%%%%%%%%%%%%%%%%%%%%%%%%%%%%%%%%%%%%%%%%%%%%%%%%%%%%%%%%%%

\section{Summary}
We have proposed a method to systematically control carrier densities at interfaces of transition-metal oxide heterostructures without introducing disorders.
By inserting non-polar layers sandwiched by polar layers, continuous control of the carrier doping into the interface can be realized.
This method enables us to control the total carrier densities per unit cell $c$ systematically up to high values of the order unity. 
We have demonstrated this charge controllability and the resultant distribution of the confined carriers at the interface by using classical and itinerant models in the Hartree mean-field theory. 

%%%%%%%%%%%%%%%%%%%%%%%%%%%%%%%%%%%%%%%%%%%%%%%%%%%%%%%%%%%%%%%%%%%%%%%%%%%%%%%    
    
\section{Acknowledgments}
The authors thank Takashi Miyake for providing us his in-house code for the LDA calculation. 
%He would also acknowledge useful advices and fruitful discussions to Shinji Watanabe, Kazuma Nakamura, Shiro Sakai, Takahiro Misawa, Youhei Yamaji, and Hiroshi Shinaoka.

\end{document}